\pdfoutput=1
\documentclass[%
 reprint,
superscriptaddress,
 amsmath,amssymb,longbibliography,
 aps,
 prb,
]{revtex4-2}%
\usepackage{hyperref}
\usepackage{multirow}
\usepackage{booktabs}
\usepackage[dvipsnames]{xcolor}
\usepackage{dsfont}
\usepackage{nicefrac}
\usepackage{bm}
\usepackage{graphicx}
\usepackage{amsmath}
\usepackage{color}
\usepackage[normalem]{ulem}
\usepackage{float}

\usepackage{tikz}

\newcommand{\mbf}[1]{\mathbf{#1}}

\newcommand{\kv}{{\bf k}}
\newcommand{\qv}{{\bf q}}

\newcommand{\io}{\tilde \mu}

\newcommand{\ial}{\tilde \mu}

\graphicspath{{./Figures/}}

\usetikzlibrary{arrows,positioning} 
\tikzset{
    photon/.style={decorate, decoration={snake}, draw=black},
    electron/.style={draw=black, postaction={decorate},
        decoration={markings,mark=at position .55 with {\arrow[draw=black,thick]{>}}}},
    gluon/.style={decorate, decoration={snake},draw=black}, 
    >=stealth',
    punkt/.style={
           rectangle,
           rounded corners,
           draw=black, very thick,
           text width=6.5em,
           minimum height=2em,
           text centered},
    pil/.style={
           ->,
           thick,
           shorten <=2pt,
           shorten >=2pt,}
}

\begin{document}

\title{Superconductivity from repulsive interactions on the kagome lattice}

\author{Astrid T. R\o mer}
\affiliation{Niels Bohr Institute, University of Copenhagen, 2200 Copenhagen, Denmark}
\affiliation{Danish Fundamental Metrology, Kogle All\'e 5, 2970 H{\o}rsholm, Denmark}

\author{Shinibali Bhattacharyya} 
\affiliation{Institut f\"ur Theoretische Physik, Goethe-Universit\"at, 60438 Frankfurt am Main, Germany}

\author{Roser Valent{\'i}} 
\affiliation{Institut f\"ur Theoretische Physik, Goethe-Universit\"at, 60438 Frankfurt am Main, Germany}

\author{Morten H. Christensen}
\affiliation{Niels Bohr Institute, University of Copenhagen, 2200 Copenhagen, Denmark}

\author{Brian M. Andersen}
\affiliation{Niels Bohr Institute, University of Copenhagen, 2200 Copenhagen, Denmark}
\email{mchriste@nbi.ku.dk}

\date{\today}

\begin{abstract}
The discovery of superconductivity in layered vanadium-based kagome metals $A$V$_3$Sb$_5$ ($A$: K, Rb, Cs) has added a new family of materials to the growing class of possible unconventional superconductors. However, the nature of the superconducting pairing in these materials remains elusive. We present a microscopic theoretical study of the leading superconducting instabilities on the kagome lattice based on spin- and charge-fluctuation mediated Cooper pairing. The applied methodology includes effects of both on-site and nearest-neighbor repulsive Coulomb interactions. Near the upper van Hove filling -- relevant for the $A$V$_3$Sb$_5$ materials -- we find a rich phase diagram with several pairing symmetries being nearly degenerate. In particular, while a substantial fraction of the phase diagram is occupied by a spin-singlet order parameter transforming as a two-dimensional irreducible representation of the point group, several nodal spin-triplet pairing states remain  competitive. We compute the band and interaction parameter-dependence of the hierarchy of the leading superconducting instabilities, and determine the detailed momentum dependence of the resulting preferred gap structures. Crucially, for moderate values of the interaction parameters, the individual pairing states depend strongly on momentum and exhibit multiple nodes on the Fermi surface. We discuss the properties of these superconducting gap structures in light of recent experimental developments of the $A$V$_3$Sb$_5$ materials.

\end{abstract}

\maketitle

\section{Introduction}

Determining the nature of the Cooper pairs that form the condensates of unconventional superconductors remains a fundamental objective of condensed matter physics. This endeavour includes a consensual experimental description of the superconducting state, in agreement with material-specific theoretical modelling. For many systems of current interest, this goal remains an outstanding research challenge, but tremendous progress has been made over the last few decades. This progress has been driven largely by an aspiration to understand pairing in, for example, the high-$T_c$ cuprates, heavy-fermion superconductors, and iron-based superconductors~\cite{Scalapino2012,Chubukovreview,Steglich2016Foundations,Leereview,HIRSCHFELD2016197,Kreisel_review,Fernandes2022Iron}. Advances have been made possible both by improvements in measurement techniques and by the discovery of entirely new classes of superconducting systems. In this respect, the recent discoveries of superconductivity in metallic layered kagome materials -- including $A$V$_3$Sb$_5$ ($A$: K, Rb, Cs)~\cite{Ortiz2019New,Ortiz2020CsV3Sb5,Ortiz2021Superconductivity}, LaRu$_3$Si$_2$~\cite{GuguchiaLaRuSi_2021}, YRu$_3$Si$_2$~\cite{Gong_2022}, and LaIr$_3$Ga$_2$~\cite{Gui2022}   -- has opened up new possibilities for exploring unconventional pairing on the kagome lattice. However, it is currently unsettled whether, e.g., the $A$V$_3$Sb$_5$ compounds actually feature unconventional pairing driven by repulsive electronic interactions, or whether a conventional phonon-driven superconductivity applies. In the former case, identifying regions of opposite sign of the superconducting gap on the Fermi surface and the possible presence of gap nodes, is important for understanding the origin of superconductivity~\cite{Shen1993Anomalously,Ding1996Angle-resolved,Kirtley1995Symmetry,Kondo2008Momentum,Reid2010Nodes,Izawa2001Angular,Chubukov2008Magnetism,Mazin2008Unconventional,Romer2015,Guterding2015,Guterding2016,Romer2020}. In the $A$V$_3$Sb$_5$ compounds, the onset of a charge-density wave (CDW) order at $T_{\rm CDW}\sim 100$~K appears to coincide with -- or precede -- the breaking of time-reversal symmetry and six-fold lattice rotational symmetry, and the appearance of a large anomalous Hall conductivity~\cite{Ortiz2019New,Yang2020Giant,Jiang2021Unconventional,Xiang2021,Chen2021Roton,Ratcliff2021,Stahl2021Temperature,Mielke2022Time-reversal,Khasanov2022Charge,Kang2022Twofold,Li2022Rotation,Nie2022Charge,Wu2022Charge,Yu2021,Guguchia2022Tunable,Guo2022Field-tuned,Hu2022Time-reversal,Jiang2022Observation}. As the CDW order is suppressed by either doping, strain, or pressure, the superconducting critical temperature, $T_c$, increases~\cite{Yu2021Unusual,Chen2021Double,Qian2021Revealing,Oey2022Fermi}. The combination of these facts points to the importance of electronic correlations in driving the ordered states in $A$V$_3$Sb$_5$. Additionally, the electronic structure exhibits multiple van Hove singularities near the Fermi level, consistent with the CDW order being driven by an instability of the Fermi surface~\cite{Ortiz2021Fermi,Kang2022Twofold,Hu2022Rich}. However, the relationship between the CDW phase and the breaking of time-reversal symmetry and lattice rotational symmetry is currently being debated~\cite{Park2021,Lin2021,Tan2021,Christensen2021,Denner2021,Feng2021Low-energy,Christensen2022Loop}.

These recent developments relate to broader studies of electronic instabilities of interacting fermions on the kagome lattice~\cite{Yu2012,Wang2013,Kiesel2013,Ferrari2022}. The electronic band structure of the kagome lattice exhibits three distinct features; a flat band, Dirac points at K, and upper and lower filling fraction van Hove points at M~\cite{Guo2009,Mazin2014,Ghimire2020Topology,Kiesel2013} (see Fig.~\ref{fig:unit_cell_and_band_structure}). The actual band structure of the $A$V$_3$Sb$_5$ materials is considerably more complex, but the Fermi surface derived from density functional theory is a hexagon-shaped Fermi contour of vanadium $3d$ character~\cite{Ortiz2019New,Tan2021}. This compares reasonably well with angle-resolved photo-emission spectroscopy measurements~\cite{Hu2022Rich,Kang2022Twofold}, and such a Fermi surface can be captured near the upper van Hove singularity of the minimal model tight-binding band~\cite{Kiesel2012,Kiesel2013}. 

The triangular network of the kagome lattice structure tends to frustrate electronic order, leading to rich phase diagrams with close competition of several ordered phases~\cite{Wen2010Interaction-driven,Kiesel2013,Wang2013}. For example, at the upper van Hove filling, the particle-hole channel is unstable to several vastly different charge- and spin-density wave ground states depending on the amplitude of the onsite and nearest-neighbor Hubbard interactions~\cite{Yu2012,Wang2013,Kiesel2013,Ferrari2022}. Similarly, the preferred particle-particle (superconducting) ground states are expected to fundamentally change depending on the interaction strength. Theoretically, earlier studies have predicted the relevance of the two-dimensional irreducible representation (irrep) $E_2$ $\{d_{x^2-y^2},d_{xy}\}$ of the $C_{6v}$ point group at the upper van Hove filling~\cite{Kiesel2012,Yu2012,Iimura2018Thermal}. More recent works, motivated by the discovery of superconductivity in the $A$V$_3$Sb$_5$ materials, have also pointed out the potential relevance of other gap symmetries for these compounds~\cite{Wu2021,Tazai2022,Wen2022,Lin2021_Multidome}.          

Experimentally, the $A$V$_3$Sb$_5$ kagome superconductors enter their superconducting phase at $T_c\sim 1-2$K~\cite{Ortiz2021Superconductivity,Ortiz2020CsV3Sb5,Yin2021Superconductivity}. However, $T_c$ may be significantly enhanced  by hydrostatic pressure, e.g, for CsV$_3$Sb$_5$ $T_c\sim 8$K at $\sim 2$~GPa~\cite{Chen2021Double,Gupta2022Two}. There is currently no experimental consensus on the detailed superconducting gap properties of these materials. Some penetration depth measurements and specific heat data on CsV$_3$Sb$_5$ point to an anisotropic gap with a finite small minimum gap~\cite{Ortiz2021Superconductivity,Duan2021,Gupta2022,Roppongi2022,Gupta2022Two}. This is consistent with a ‘U’-shaped conductance behavior near zero-bias from scanning tunneling microscopy (STM) measurements~\cite{Xu_PRL_2021}. The lack of sign-changes in the gap function, i.e. nodes on the Fermi surface, is consistent with the absence of in-gap states near non-magnetic impurities~\cite{Xu_PRL_2021}. Furthermore, $s$-wave spin-singlet order has been proposed based on an observed Knight shift suppression and the existence of a Hebel-Slichter nuclear magnetic resonance (NMR) peak~\cite{Mu_2021}. On the other hand, several STM measurements have also reported ‘V’-shaped STM conductance spectra~\cite{Chen2021Roton,Liang2021}, and thermal conductivity data has been interpreted in favor of nodal superconductivity~\cite{Zhao_2021}. Similarly, recent muon spin spectroscopy experiments on RbV$_3$Sb$_5$ and KV$_3$Sb$_5$ samples report nodal gaps at ambient pressure~\cite{Guguchia2022Tunable}, and a pressure-tuned transition to nodeless order with additional evidence for spontaneous time-reversal symmetry breaking (TRSB) setting in at $T_c$ for high pressures $\sim 2$~GPa~\cite{Mielke2022Time-reversal,Guguchia2022Tunable}. Finally, some experiments have pointed to other kinds of unusual superconducting order in the $A$V$_3$Sb$_5$ systems~\cite{Chen2021Roton,Liang2021,Wang_triplet,Ge2022Discovery}. For example, Josephson junction studies led to the proposal of spin-triplet supercurrents in K$_{1-x}$V$_3$Sb$_5$~\cite{Wang_triplet}, and STM measurements on CsV$_3$Sb$_5$ was interpreted in terms of pair-density wave order~\cite{Chen2021Roton}. 

These recent developments motivate further comprehensive microscopic studies of the preferred superconducting pairing symmetry in kagome lattices. Here, based on a minimal kagome band structure we map out the phase diagram of the superconducting order near the two van Hove fillings. The pairing is assumed to be mediated by spin- and charge-fluctuations derived from the kagome Hubbard model including onsite ($U$) and nearest-neighbor ($V$) interactions. Such a fluctuation-driven mechanism for pairing in Hubbard-like models has recently gained renewed support~\cite{Maier2007_1,Maier2007_2,Dong2022,Romer2020}. As a function of interaction parameters, we find that the phase diagram near the upper van Hove filling exhibits several competing symmetry-distinct pairing channels, including possibilities of chiral TRSB singlet and triplet order. 
From this perspective, superconductivity appears rather frustrated at this filling fraction, and relatively small changes to the model can tip the balance in the hierarchy of the leading superconducting solutions. This is also reflected in a large number of nodes present on the Fermi surface for most of the phase diagram. We discuss these findings in light of the recent experimental and theoretical developments of superconductivity in the $A$V$_3$Sb$_5$ materials. 

\section{Model and Methods}

\begin{figure}[t]
 \centering
   	\includegraphics[width=0.9\columnwidth]{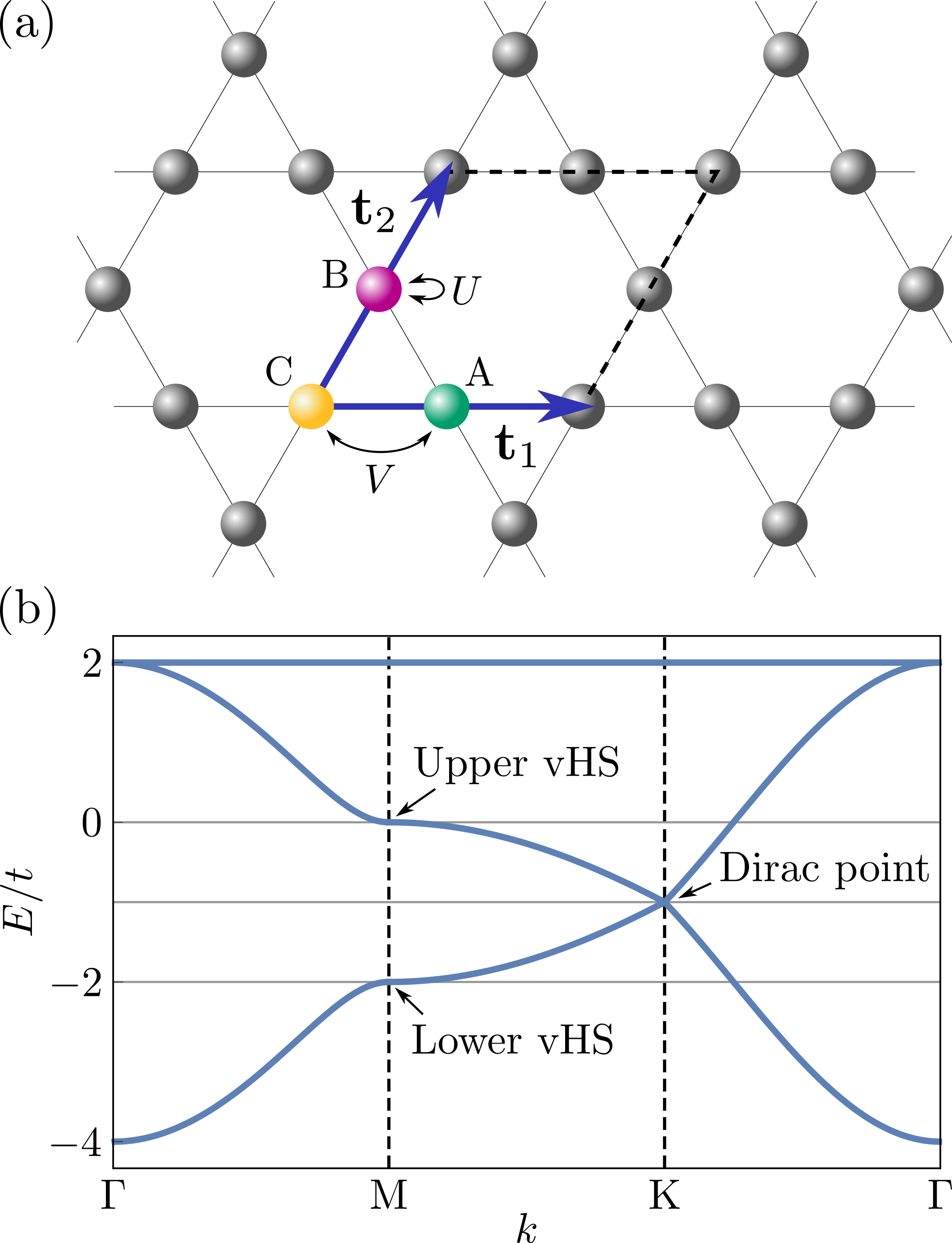}
\caption{(a) Illustration of the simple kagome lattice with the unit cell denoted by the dashed line. The three atoms in the unit cell are denoted $A$, $B$, and $C$ and are highlighted in color. The two lattice vectors, $\mathbf{t}_1$ and $\mathbf{t}_2$, are indicated in blue. $U$ and $V$ denote the onsite and nearest-neighbor Coulomb interactions, respectively. (b) Band structure of the simple kagome lattice, Eq.~\eqref{eq:Hgauge} for $t=1$, with upper and lower vHS at the $M$ point of the Brillouin zone and a Dirac point at the K point.}
\label{fig:unit_cell_and_band_structure}
\end{figure}

The simple kagome lattice has three atoms in the unit cell and is shown in Fig.~\ref{fig:unit_cell_and_band_structure}(a). The two lattice vectors, $\mathbf{t}_1$ and $\mathbf{t}_2$, are shown in blue and are given by
\begin{equation}
    \mathbf{t}_1 = \begin{pmatrix}
    a & 0
    \end{pmatrix}\,, \qquad 
    \mathbf{t}_2 = \begin{pmatrix}
    \frac{a}{2} & \frac{\sqrt{3}a}{2}
    \end{pmatrix}\,.
\end{equation}
We will set $a=1$ in what follows, thus fixing a length scale. The unit cell contains three sites, labeled $A$, $B$, and $C$, and the kinetic part of the normal-state Hamiltonian is given by
\begin{eqnarray}
 \mathcal{H}_0 = \sum_{\kv s} \Psi_{\kv,s}^\dagger H_0(\kv)\Psi_{\kv,s}\,,
 \label{eq:H0}
\end{eqnarray}
with $\Psi_{\kv,s}^\dagger=(c_{\kv A s }^\dagger,c_{\kv B s }^\dagger,c_{\kv C s }^\dagger)$ and
\begin{equation}
H_0(\kv)=\begin{pmatrix}
 -\mu & -t(1+e^{2i k_3}) & -t(1+e^{-2i k_1}) \\
-t(1+e^{-2i k_3}) & -\mu & -t(1+e^{-2i k_2}) \\
-t(1+e^{2i k_1}) & -t(1+e^{2i k_2}) & -\mu 
   \end{pmatrix}
   \label{eq:Hgauge}
\end{equation}
with $k_i=\kv \cdot {\bf a}_i$. Here, $\mathbf{a}_i$ denotes one of the three intra-unit cell vectors
\begin{equation}
    \mathbf{a}_1 = \frac{1}{2}\mathbf{t}_1\,, \qquad \mathbf{a}_2=\frac{1}{2}\mathbf{t}_2\,, \qquad \mathbf{a}_3 = \mathbf{a}_2 - \mathbf{a}_1\,.
\end{equation}
The operator $c^\dagger_{\kv \mu s}$ ($c_{\kv \mu s}$) denotes the creation (annihilation) operator of an electron with momentum $\kv$, spin $s$, and sublattice flavor $\mu= A,B,C$. The nearest-neighbor hopping $t=1$ sets the energy scale.

The bare onsite interaction Hamiltonian in real space is given by
\begin{equation}
    \mathcal{H}_{\rm U}=U \sum_{i \mu s} n_{\mu,i, s}n_{ \mu,i,\overline{s}}\,, \label{eq:Hu}
\end{equation}
while the nearest-neighbor repulsion reads
\begin{equation}
\mathcal{H}_{\rm V} = \sum_{\substack{ij \\ \mu s s'}} n_{\mu, i, s} n_{\overline{\mu},j,s'}\,,\label{eq:Hv}
\end{equation}
where $n_{\mu,i,s}=c^\dagger_{\mu,i,s}c_{\mu,i,s}$ and $\overline s=-s$. Here, $\overline{\mu}$ denotes the two sublattice indices distinct from $\mu$, e.g., for $\mu=A$, $\overline{\mu}=B, C$. $i$ and $j$ denote identical or neighboring unit cells and $U,V>0$ correspond to the onsite and nearest-neighbor interaction, respectively. 

We parameterize the Fermi surface by $300$ wavevector points and extract the superconducting instabilities from the linearized BCS gap equation
\begin{equation}
  -\frac{\sqrt{3}}{2(2\pi)^2}\oint_{\rm FS} d \kv_f^\prime \frac{1}{|v(\kv_f^\prime)|} \Gamma_{s/t}(\kv_f,\kv_f^\prime)\Delta(\kv_f^\prime)=\lambda \Delta(\kv_f)\,,
  \label{eq:LGE}
\end{equation}
where the pairing kernel $\Gamma_{s/t}(\kv_f,\kv_f^\prime)$ in the singlet ($s$) and triplet ($t$) channel is given by
\begin{equation}
\Gamma_{s/t}(n_1\kv,n_2\kv')=
\sum_{\{s\}} 
[\mathcal{V}(n_1\kv,n_2\kv')]^{s ~\overline s}_{\overline s ~s}~\mp~
[\mathcal{V}(n_1\kv,n_2\kv')]^{s ~s}_{\overline s ~\overline s}\,,
\label{eq:GammaLL}
\end{equation}
and $[\mathcal{V}(n_1\kv,n_2\kv')]^{s_1 s_2}_{s_3 s_4}$ denotes the spin-dependent effective interaction projected to band-space with band labels $n_1$ and $n_2$.
In Eq.~\eqref{eq:LGE}, $\kv_f$ denotes a wavevector on the Fermi surface and the band label is suppressed, since it is uniquely determined by $\kv_f$. The Fermi velocity is denoted by $v(\kv_f)$ and the leading superconducting instability is given by the gap function, $\Delta(\kv_f)$, with the largest eigenvalue $\lambda$~\cite{Romer2015}.
Only intraband Cooper pairing is included. Below we provide further details of the computation of the spin- and charge-fluctuation driven pairing kernel $\Gamma_{s/t}(\kv_f,\kv_f^\prime)$. Readers uninterested in further technical details can skip to Sec.~\ref{sec:results}. 

The effective electron-electron interaction in the Cooper channel is derived from
onsite and nearest-neighbor Coulomb repulsion as stated in Eqs.~\eqref{eq:Hu} and \eqref{eq:Hv}.
Restricting attention to the Cooper channel, the bare interaction, $\mathcal{H}^{\rm SC}_{\rm U+V}$, can be written as 
\begin{equation}
 \mathcal{H}^{\rm SC}_{\rm U+V}=\sum_{\kv \kv' \tilde{\mu}}
 [W_0(\kv-\kv')]^{\ial_1\ial_2}_{\ial_3\ial_4}   
 c^\dagger_{\ial_1, \kv}
 c^\dagger_{\ial_3, -\kv}
  c_{\ial_2, -\kv'}
  c_{\ial_4, \kv'}\,,
\end{equation}
where we introduced the combined index $\tilde{\mu}=(\mu,s)$. The elements of $[W_0(\mathbf{q})]^{\ial_1\ial_2}_{\ial_3\ial_4}$ are defined by 
\begin{equation}
\begin{split}
    [W_0(\qv)]^{\mu s,\mu \overline{s}}_{\mu \overline{s},\mu s} &= U\,, \\
    [W_0(\qv)]^{\mu s,\mu s}_{\mu \overline{s},\mu \overline{s}} &= -U, 
\end{split}\label{eq:U_Cooper}
\end{equation}
and
\begin{equation}
\begin{split}
[W_0(\qv)]^{A s~,B s'}_{B s',A s} &= V(1+e^{-2i\qv\cdot {\bf a}_3})\,, \\
\left[W_0(\qv)\right]^{A s~,C s'}_{C s',A s} &= V(1+e^{2i\qv\cdot {\bf a}_1})\,,\\
 \left[W_0(\qv)\right]^{B s~,C s'}_{C s',B s} &= V(1+e^{2i\qv\cdot {\bf a}_2})\,,
\end{split}\label{eq:V_Cooper}
\end{equation}
with $[W_0(\qv)]^{B s~,A s'}_{A s',B s}={\rm conj}([W_0(\qv)]^{A s~,B s'}_{B s',A s})$.

Effective attractive interactions result from higher order processes which are derived from all bubble and ladder diagrams summed to infinite order~\cite{Romer2015,Romer2019}. This gives rise to a compact formulation of the pairing interaction -- the random phase approximation (RPA) -- which leads to pairing mediated by spin and charge fluctuations. Within RPA, we can write the total interaction in the Cooper channel as
\begin{equation}
    \mathcal{H}^{\rm SC}_{\rm RPA} = \sum_{\mbf{k}\mbf{k'}\tilde{\mu}} [\mathcal{V}(\mbf{k},\mbf{k'})]^{\tilde{\mu}_1\tilde{\mu}_2}_{\tilde{\mu}_3\tilde{\mu}_4}c^{\dagger}_{\tilde{\mu}_1,\mbf{k}}c^{\dagger}_{\tilde{\mu}_3,-\mbf{k}}c_{\tilde{\mu}_2,-\mbf{k'}}c_{\tilde{\mu}_4,\mbf{k'}}\,,
\end{equation}
where $[\mathcal{V}(\mbf{k},\mbf{k'})]^{\tilde{\mu}_1\tilde{\mu}_2}_{\tilde{\mu}_3\tilde{\mu}_4}$ is a sum of the bare and renormalized interactions
\begin{align}
    [\mathcal{V}(\mbf{k},\mbf{k'})]^{\tilde{\mu}_1\tilde{\mu}_2}_{\tilde{\mu}_3\tilde{\mu}_4} &= [W_0(\mbf{k}-\mbf{k'})]^{\tilde{\mu}_1\tilde{\mu}_2}_{\tilde{\mu}_3\tilde{\mu}_4} - [W_0(\mbf{k}+\mbf{k'})]^{\tilde{\mu}_1\tilde{\mu}_4}_{\tilde{\mu}_3\tilde{\mu}_2} \nonumber \\ &+ [V_{\rm fluc}(\mbf{k},\mbf{k'})]^{\tilde{\mu}_1\tilde{\mu}_2}_{\tilde{\mu}_3\tilde{\mu}_4}\,.
\end{align}
The term $[V_{\rm fluc}(\mbf{k},\mbf{k'})]^{\tilde{\mu}_1\tilde{\mu}_2}_{\tilde{\mu}_3\tilde{\mu}_4}$ arises due to the RPA resummation of bubble and ladder diagrams and is given by
\begin{widetext}
\begin{align}
[V_{\rm fluc}(\kv,\kv')]^{\io_1 \io_2}_{\io_3 \io_4} &= \sum_{\delta \delta'} e^{-i\delta\kv}e^{i\delta'\kv'}\Big[
[W(\kv+\kv',\delta)]\Big[[1-\chi_0W]^{-1}\chi_0\Big](\kv+\kv',\delta,\delta')
[W(\kv+\kv',\delta')]\Big]^{\io_1 \io_2}_{\io_3 \io_4}\nonumber \\
& - \sum_{\delta \delta'}e^{-i\delta\kv}e^{-i\delta'\kv'}\Big[
[W(\kv-\kv',\delta)]\Big[[1-\chi_0W]^{-1}\chi_0\Big](\kv-\kv',\delta,\delta')
[W(\kv-\kv',\delta')]\Big]^{\io_1 \io_4}_{\io_3 \io_2}\,.
\label{eq:Veff_fluc}
\end{align}
\end{widetext}
Here, we have introduced a generalized static susceptibility defined by
\begin{align}
\Big[\chi_0(\qv,\delta,\delta')\Big]^{\mu_1s ~ \mu_2s'}_{\mu_3s'~ \mu_4s}
&=\frac{1}{N}\sum_{\substack{\kv \\ n_1 n_2}} e^{i\kv(\delta-\delta')} M_{n_1 n_2}^{\mu_1\mu_2\mu_3\mu_4}(\mbf{k},\mbf{q}) \nonumber \\ &\times
 \frac{f(\xi_{\kv ,n_2 })-f(\xi_{\kv-\qv, n_1 })}{\xi_{\kv-\qv, n_1}-\xi_{\kv, n_2 }}\,,
\label{eq:BareSus}
\end{align}
with the matrix element
\begin{equation}
    M_{n_1 n_2}^{\mu_1\mu_2\mu_3\mu_4}(\mbf{k},\mbf{q}) = [u_{n_1}^{\mu_1}(\kv-\qv)]^*  [u_{n_2}^{\mu_3} (\kv)]^* u_{n_2}^{\mu_2} (\kv)u_{n_1}^{\mu_4} (\kv-\qv)\,,
\end{equation}
where $u_{n}^{\mu}(\kv)$ is an eigenvector of $H_0(\mbf{k})$ giving the transformation from sublattice space to band space. $\xi_{\mbf{k},n}$ is an eigenvalue of $H_0(\mbf{k})$ and $f(\xi)=(e^{\xi/k_{\rm B}T}+1)^{-1}$ is the Fermi-Dirac distribution. The bare susceptibility, $\chi_0$, and the interaction matrices, $W$, depend on four sublattice and spin indices, $\tilde \mu_i=(\mu_i,s_i)$, as well as the vectors $\delta \in \{{\bf 0},{\bf t}_1,-{\bf t}_1,{\bf t}_2,-{\bf t}_2,{\bf t}_2-{\bf t}_1,-{\bf t}_2+{\bf t}_1\}$. For instance, $W(\mbf{q},\delta=0)=W_0(\mbf{q})$, whereas $W(\mbf{q},\delta=\mbf{t}_1)=V$ corresponds to the interaction between nearest-neighbor A and C sites with no phase factor appended. With the three sublattices, two spin degrees of freedom, and seven different values of $\delta$, this results in a $(3\cdot 2)^2 \cdot 7 \times (3\cdot 2)^2 \cdot 7$-dimensional matrix structure. In Eq.~\eqref{eq:Veff_fluc}, the matrix products are implicit and internal indices, such as $\io$ and $\delta$, are suppressed. The ladder diagrams involves book-keeping of phase factors $e^{2i\qv\cdot {\bf a}_i}$, where in general, the momentum $\qv$ is an internal momentum of the ladder diagram~\cite{Romer2021}. This is ensured by introducing the phase factors $e^{i\kv (\delta-\delta')}$ in Eq.~\eqref{eq:BareSus}. We obtain the effective interaction, $[\mathcal{V}(n_1\kv,n_2\kv')]^{s_1 ~s_2}_{s_3 ~ s_4}$, of Eq.~\eqref{eq:GammaLL} by projecting $[\mathcal{V}(\kv,\kv')]^{\ial_1\ial_2}_{\ial_3\ial_4}$ onto band space using the eigenvectors $u_{n}^{\mu}(\kv)$. The susceptibilities are computed on a $120\times 120 $ grid at a temperature $k_{\rm B}T=0.01$, which is representative of the low-temperature limit. We have checked that the results are robust to moderate changes in temperature.

Despite the fact that $s$ and $s'$ denote spin indices in Eq.~\eqref{eq:BareSus}, the bare susceptibility is independent of spin. The indices are included for a systematic treatment of the spin-dependent vertices in the diagrammatic expansion, Eq.~\eqref{eq:Veff_fluc}. This also allows us to define the RPA charge- and spin-susceptibility matrices given by
\begin{equation}
    [\chi_{\rm sp/ch}(\mbf{q})]^{\mu_1 ~ \mu_2 }_{\mu_3  ~ \mu_4 } =\sum_{s} \left(
    [\chi(\mbf{q})]^{\mu_1 s ~ \mu_2 s}_{\mu_3 s ~ \mu_4 s}
    \mp
    [\chi(\mbf{q})]^{\mu_1 s ~ \mu_2 s}_{\mu_3 \overline s ~ \mu_4 \overline s} \right). \label{eq:susc_def}
\end{equation}
In section ~\ref{sec:Results_susc} below, we explore how the leading eigenvalues, $\lambda_{\rm sp/ch}(\mbf{q})$, resulting from the diagonalization of the susceptibility matrices in Eq.~\eqref{eq:susc_def}, evolve upon increasing interactions $U$ and $V$.
The emergent structures and their relation to the preferred superconducting gap structure are discussed in Secs.~\ref{sec:sc_upper_vh} and \ref{sec:sc_lower_vh}.

\begin{figure*}[t]
 \centering
   	\includegraphics[angle=0,width=0.95\linewidth]{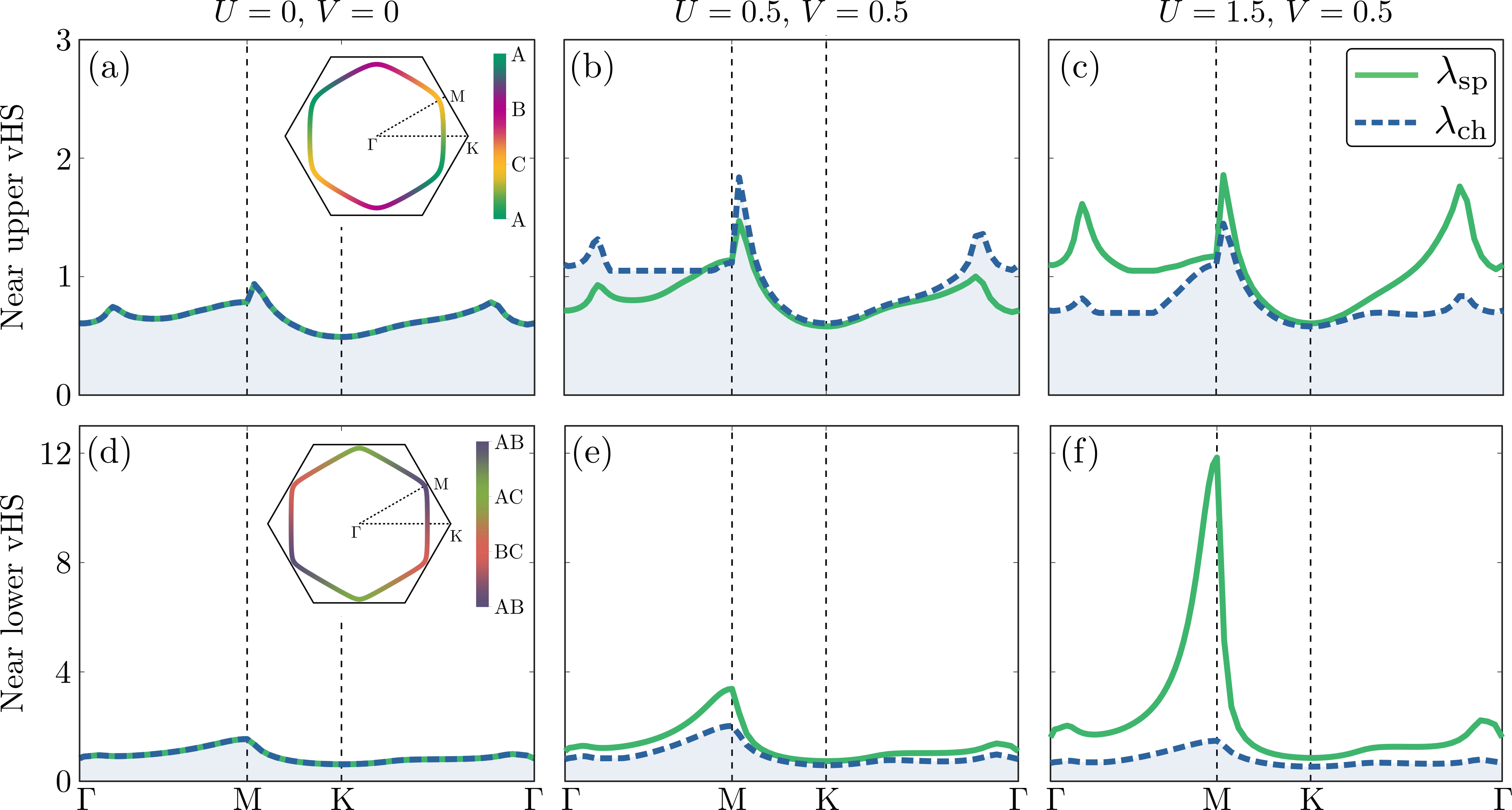}
\caption{Leading eigenvalue of the particle-hole susceptibility projected onto the spin and charge channels, as given by Eq.~\eqref{eq:susc_def}. (a)--(c) display the susceptibilities in the vicinity of the upper van Hove filling, $\mu=0.08$, corresponding to electron filling of $n\approx 5/12+0.02$. In (a), we show the bare susceptibility eigenvalues while the inset depicts the Fermi surface with dominant sublattice contribution displayed by the color code. The dotted line highlights the high-symmetry path along which the eigenvalues are plotted. In (b) and (c) we show the RPA susceptibility eigenvalues for $V=0.5$, and $U=0.5$ and $U=1.5$, respectively. The spin- and charge sectors are of comparable amplitudes throughout the BZ. (d)--(f) Same as (a)--(c), but near the lower van Hove filling with $\mu=-2.02$, corresponding to $n \approx 1/4-0.006$. In this case, the spin susceptibility strongly dominates in the large-$U$ regime.}
\label{fig:chiRPA}
\end{figure*}

\section{Results}\label{sec:results}

As illustrated in Fig.~\ref{fig:unit_cell_and_band_structure}(b), the band structure of the simple one-orbital kagome lattice features a flat band, Dirac points at $K$, and two different van Hove points at the $M$ points of the Brillouin zone (BZ)~\cite{Guo2009,Ghimire2020Topology}. The upper van Hove point is located at an electron filling of $5/12$ while the lower one is at a filling of $1/4$. In the following, we focus on solutions to the linearized gap equation, Eq.~\eqref{eq:LGE}, near the two van Hove fillings. While the $A$V$_3$Sb$_5$ materials feature a fermiology similar to the band structure near the upper van Hove point~\cite{Kang2022Twofold}, it is instructive to compare the results between the upper and lower van Hove singularities. Despite the two fillings exhibiting the same Fermi surface topology, the nature of the eigenstates at the Fermi level is very different, due to the distribution of the sublattice weight on the Fermi surface~\cite{Kiesel2012}. The pairing interaction is primarily affected by the spin and charge susceptibilities, as seen from Eq.~\eqref{eq:Veff_fluc}. Hence, the structure of the susceptibilities play a crucial role in determining the symmetry of the superconducting instability. Therefore, we consider first the variation in the susceptibilities between the upper and lower van Hove fillings.
\begin{table}[t!]
    \includegraphics[width=0.98\columnwidth]{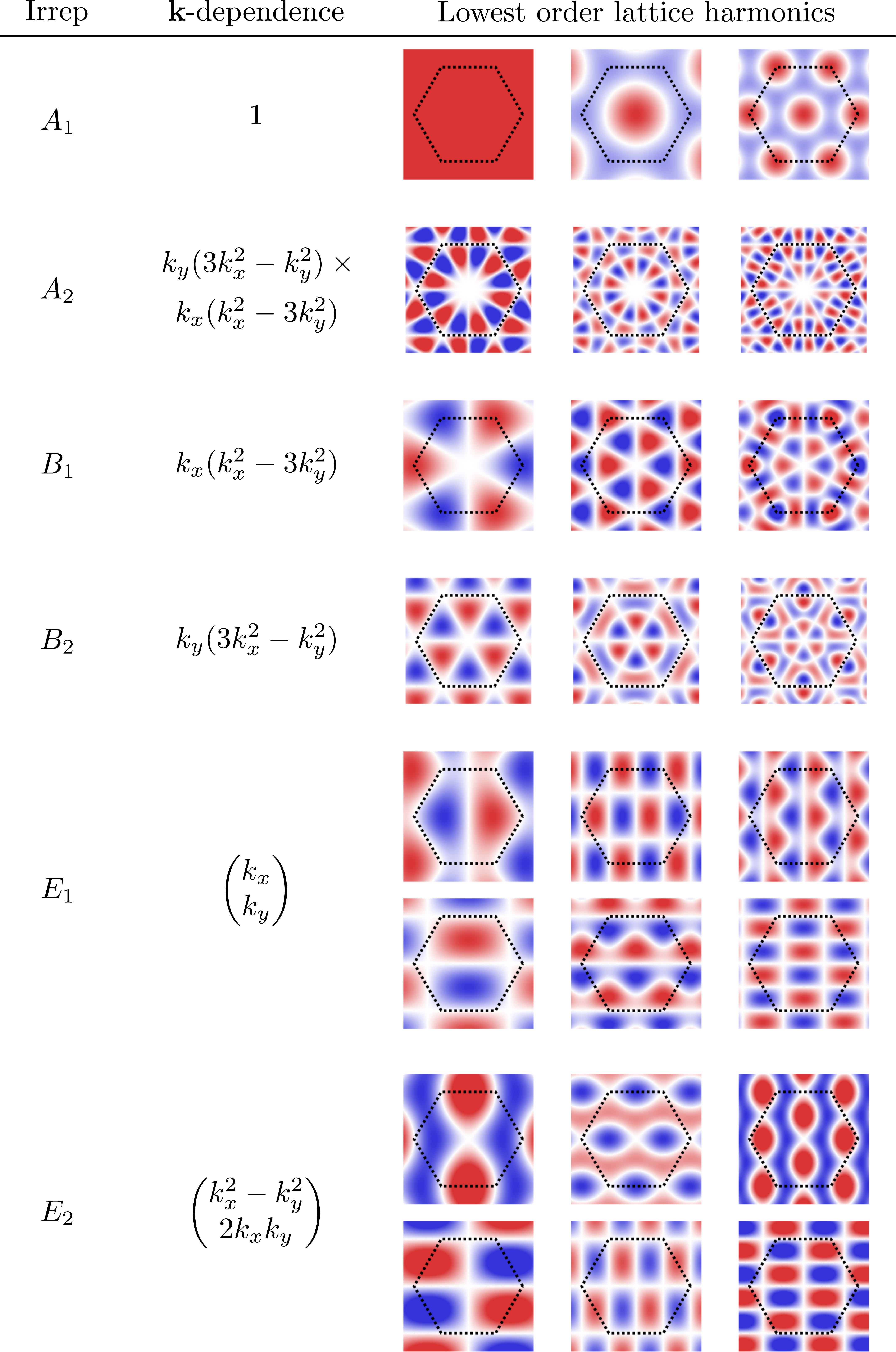}
    \caption{Irreps of $C_{6v}$, their lowest-order functional dependence on $k_x$ and $k_y$, and the three lowest order lattice harmonics. The BZ is shown by the dashed outline and we note that the $A_2$ and $B_2$ irreps have protected nodes along the BZ boundary.}
    \label{tab:irrep_table}
\end{table}

\subsection{Susceptibilities near the van Hove fillings}\label{sec:Results_susc}

The RPA charge- and spin-susceptibility matrices are defined in Eq.~\eqref{eq:susc_def}. In Figs.~\ref{fig:chiRPA}(a)--(c) and Figs.~\ref{fig:chiRPA}(d)--(f) we display the momentum structure of the leading eigenvalue, $\lambda_{\rm sp/ch}(\mbf{q})$, of the susceptibility matrices $ [\chi_{\rm sp/ch}(\mbf{q})]^{\mu_1 ~ \mu_2 }_{\mu_3  ~ \mu_4 }$ near the upper and lower van Hove filling, respectively, along the high-symmetry path indicated in the inset in Fig.~\ref{fig:chiRPA}(a). The variation in sublattice distribution, shown in the insets of Figs.~\ref{fig:chiRPA}(a) and \ref{fig:chiRPA}(d), has a clear impact on the susceptibilities and the associated eigenvalues. Figures~\ref{fig:chiRPA}(a) and \ref{fig:chiRPA}(d) depict the bare susceptibilities, where there is no distinction between the spin and charge channels. In Figs.~\ref{fig:chiRPA}(b) and \ref{fig:chiRPA}(c) we show the RPA susceptibility eigenvalues near the upper van Hove filling for increasing values of $U$ and fixed $V=0.5$. As expected, the on-site interaction $U$ promotes spin order, while $V$ promotes the formation of charge order. Similarly, in Figs.~\ref{fig:chiRPA}(e) and \ref{fig:chiRPA}(f) we show the RPA susceptibility eigenvalues near the lower van Hove filling for the same values of $U$ and $V$. Even though the two cases share Fermi surface topology, the distinct sublattice weights lead to a very different momentum structure of the susceptibilities. Near the lower van Hove point, the interactions favor the formation of spin order, indicated by the dominant peak at M. In contrast, the eigenvalues near the upper van Hove filling are significantly reduced and exhibit several peaks of comparable amplitude, as seen in Figs.~\ref{fig:chiRPA}(b) and (c).

Despite these important differences, the tendency towards spin order driven by the peak at $M$ in the regime of $U>V$ is a common denominator between the upper and lower van Hove fillings. Further investigations reveal that the spin order is of a mixed-sublattice character. Similarly, the structure of the charge peak found in the regime where $V$ dominates indicates that bond (inter-site) fluctuations are prominent.
On the other hand, the small-$\bf{q}$ peak visible in the spin channel [green curve in Fig.~\ref{fig:chiRPA}(c)] is dominated by intra-site fluctuations. 
The tendency towards either several competing orders near the upper van Hove point or one dominant order near the lower van Hove point has important consequences for the preferred pairing states as discussed in detail below.

\subsection{Leading superconducting solutions near the upper van Hove filling}
\label{sec:sc_upper_vh}

The allowed pairing symmetries of the simple two-dimensional kagome lattice can be classified by the irreps of the $C_{6v}$ point group~\cite{Wen2022}. These are denoted by $A_1$, $A_2$, $B_1$, and $B_2$ for the one-dimensional irreps, and $E_1$ and $E_2$ for the two-dimensional irreps. These are summarized in Table~\ref{tab:irrep_table}. In this table, we also include the three lowest order lattice harmonics of each irrep. Of these irreps, $A_1$, $A_2$, and $E_2$ correspond to spin-singlet solutions, whereas $B_1$, $B_2$, and $E_1$ solutions are spin-triplet in nature. As spin-orbit coupling is neglected in our model, these refer to true spin states. The two one-dimensional singlet solutions, $A_1$ and $A_2$, feature zero and 12 symmetry-imposed nodes, respectively, as seen from Table~\ref{tab:irrep_table}. Additionally, the $A_2$ solution has a node along the BZ boundary. The lowest-order lattice harmonic of the $E_2$ spin-singlet solution is the $\{d_{x^2-y^2},d_{xy} \}$ doublet, each with four nodes. The two one-dimensional spin-triplet solutions, $B_1$ and $B_2$, exhibit six symmetry-protected nodes each. The nodes of $B_1$ are along the $\Gamma - {\rm M}$-path in the BZ while $B_2$ has nodes along the $\Gamma - {\rm K}$-path, i.e., between the corners of the BZ. The $B_2$ solution also has a node along the BZ boundary. Finally, the lowest-order lattice harmonic of the two-dimensional spin-triplet solution, $E_1$, corresponds to the $\{p_x, p_y\}$-state familiar from, e.g., the study of topological superconductors~\cite{Read2000Paired}. In addition to the nodes imposed by symmetry, the gap structures we discuss below exhibit accidental nodes. These arise from the superposition of the lattice harmonics shown in Table~\ref{tab:irrep_table}, alongside the higher-order contributions. In general, these accidental nodes depend on the details of the microscopic model~\cite{Romer2021}.
\begin{figure*}
 \centering
 \includegraphics[width=\textwidth]{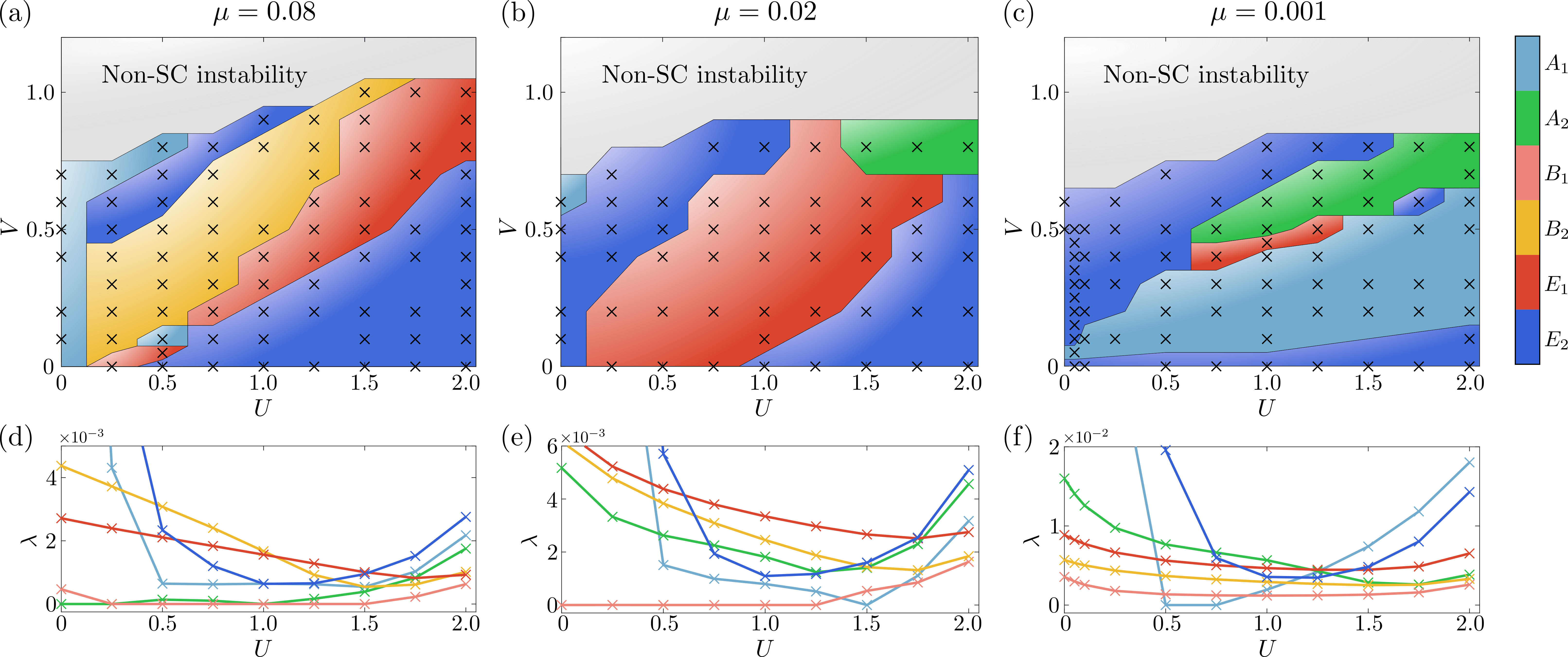}
\caption{(a)--(c) Phase diagrams showing the leading superconducting pairing symmetry as a function of onsite interaction, $U$, and  NN interaction, $V$. The three panels (a)--(c) correspond to slightly different fillings all near the upper van Hove filling: (a) $\mu=0.08$, corresponding to an electron density of $n\approx 5/12+0.02$; (b) $\mu=0.02$, $n \approx 5/12 + 0.006$; (c) $\mu=0.001$, $n \approx 5/12+0.0007$. Crosses indicate the parameter values for which the linearized gap equation was solved; the phase boundaries have been determined from these by linear interpolation. The gray color indicates regions in which the spin or charge susceptibility has diverged, thus leading to a non-SC instability. Comparison of panels (a)--(c) highlights the sensitivity of the pairing solutions to relatively small Fermi surface changes. For each case of filling there is also significant dependence on the interaction parameters $U$ and $V$. This is further illustrated by the plots in panels (d)--(f) showing the leading eigenvalue associated with each distinct irreducible representation, thus also including subleading solutions, along a $U$-cut with constant $V=0.5$.   
}
\label{fig:phasediagrams}
\end{figure*}

\begin{figure*}[tb]
 \centering
   	\includegraphics[width=\textwidth]{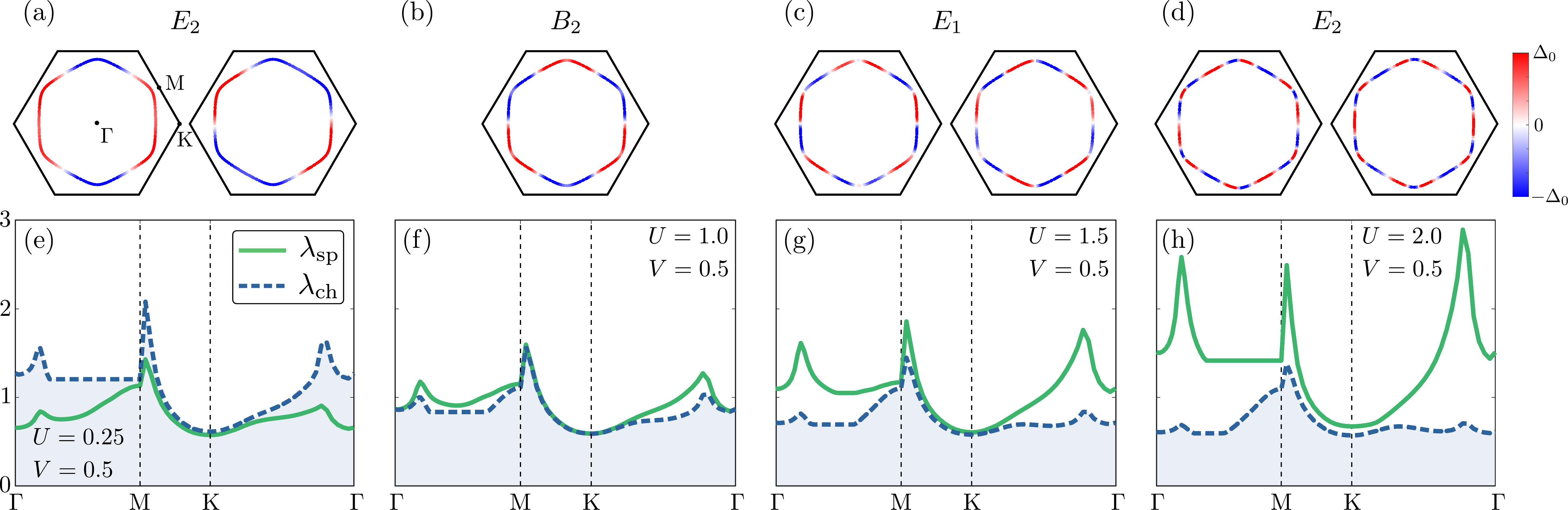}
\caption{Superconducting gap structures and susceptibility eigenvalues near the upper van Hove filling for increasing values of $U$. The gap structures correspond to the different solutions obtained along the line with $V=0.5$ in Fig.~\ref{fig:phasediagrams}(a) (we omit the highy unusual case with $U=0$ but $V>0$.) For $U=0.25$ we obtain the leading solution shown in (a). This corresponds to the two-dimensional spin-singlet $E_2$ irrep which, in this case, is dominated by the lowest-order lattice harmonics $\{ d_{x^2-y^2}, d_{xy}\}$, each with four nodes on the Fermi surface. In (e), the associated susceptibility eigenvalues are shown. The charge bond-order susceptibility are slightly dominant near $M$. Between $U=0.5$ and $U=1.5$ we obtain two distinct triplet solutions. In (b) the $U=1.0$ case is shown. This is the one-dimensional $B_2$ irrep with an $f$-wave structure (see Table~\ref{tab:irrep_table}). In (f), the spin and charge  susceptibilities are nearly identical. For $U=1.5$, shown in (c), we instead obtain a highly nodal variety of the two-dimensional $E_1$ irrep, with substantial contributions from lattice harmonics beyond the lowest order. The associated susceptibility eigenvalues in (g) are very similar to (f), although the spin fluctuations are enhanced. The $U=2.0$ case is shown in (d). Here, we also find that the $E_2$ irrep is favored, although these solutions are highly nodal, exhibiting 20 sign-changes around the Fermi surface, and lattice harmonics beyond the leading order are required to describe such a gap structure. In this case, as shown in (h) the susceptibilities are dominated by spin fluctuations but several peaks with comparable amplitudes are present.}
\label{fig:gapsuppervHd}
\end{figure*}

The preferred pairing state is determined by a complex interplay between the electronic structure and the interaction parameters~\cite{Romer2019}. This is evidenced in Fig.~\ref{fig:phasediagrams} which displays the leading pairing symmetry at three slightly different chemical potentials near the upper van Hove filling as a function of $U$ and $V$. For example, Fig.~\ref{fig:phasediagrams}(a) where $\mu=-0.08$ ($n\approx 5/12+0.02$), the $U-V$ phase diagram is seen to contain significant regions of $E_2$ spin-singlet order, $B_{2}$ and $E_1$ spin-triplet pairing symmetries, and even $A_{1}$ spin-singlet pairing. The latter fully-gapped state is present in the somewhat unphysical regime of finite $V$ but $U\simeq 0$. It emerges from a constructive interference-like effect whereby the effective pairing generated from nearest-neighbor $V$ processes produce a substantial attractive onsite pairing potential. We will not discuss this state further here. 

As seen from Fig.~\ref{fig:phasediagrams}, superconductivity near the upper van Hove filling is ``frustrated'' in the sense that several symmetry-distinct pairing channels are closely competing. This is evident both from the large variation in the phase diagrams in Figs.~\ref{fig:phasediagrams}(a)--(c), and the cuts at $V=0.5$ showing the subleading eigenvalues displayed in Figs.~\ref{fig:phasediagrams}(d)--(f). As seen, the eigenvalues cross and no particular solution ``splits off'' as in, e.g., the cuprates ($d_{x^2-y^2}$-wave) or iron-pnictides ($s_\pm$). This near-degeneracy implies that relatively small parameter changes may fundamentally alter both the momentum- and spin-structure of the Cooper pairs. The abrupt changes in the slopes of the eigenvalue curves, e.g., for the $A_1$ solution in Fig.~\ref{fig:phasediagrams}(f), reflect a change in gap structure in which the contributing lattice harmonics are modified. Similar gap structure changes are responsible for the non-monotonic eigenvalue evolution in the other symmetry channels seen in Fig.~\ref{fig:phasediagrams}(d)--(f).

From Fig.~\ref{fig:phasediagrams} it is further evident that both the $E_1$ and $E_2$ irreps are prominent candidates as the leading instability in large parts of parameter space, particularly in the physically more reasonable regime where $U \gtrsim 2V$. These solutions (irreps) are both two-dimensional. This implies that, upon the onset of a superconducting order transforming as either of these two irreps, an additional symmetry must be broken. This symmetry-breaking occurs as a result of the system choosing a specific value for the relative phase, $\varphi$, between the two components of the order parameter. In general, a relative phase of either $\varphi=0$ or $\varphi=\frac{\pi}{2}$ minimizes the free energy, corresponding to either a real or an imaginary superposition of the two components~\cite{Sigrist1991Phenomenological}. The real superposition breaks the six-fold rotational symmetry of the lattice, while the imaginary one breaks time-reversal symmetry. Under a number of simplifying assumptions, the TRSB variant is the lower-energy state~\cite{Nandkishore2012Chiral}. This is also consistent with the fact that this state has fewer nodes than the real counterpart and should therefore be associated with an enhanced condensation energy. Below, we will therefore focus on the TRSB variants of the two-dimensional irreps, rather than the ones breaking rotational symmetry. The former are chiral and thus lead to a topologically non-trivial superconducting state. Finally, we note that several distinct irreps may condense simultaneously near the phase boundaries in Fig.~\ref{fig:phasediagrams}. As a result, other exotic TRSB superconducting order parameters may also emerge in these particular regions.

\begin{figure}
 \centering
   	\includegraphics[width=\columnwidth]{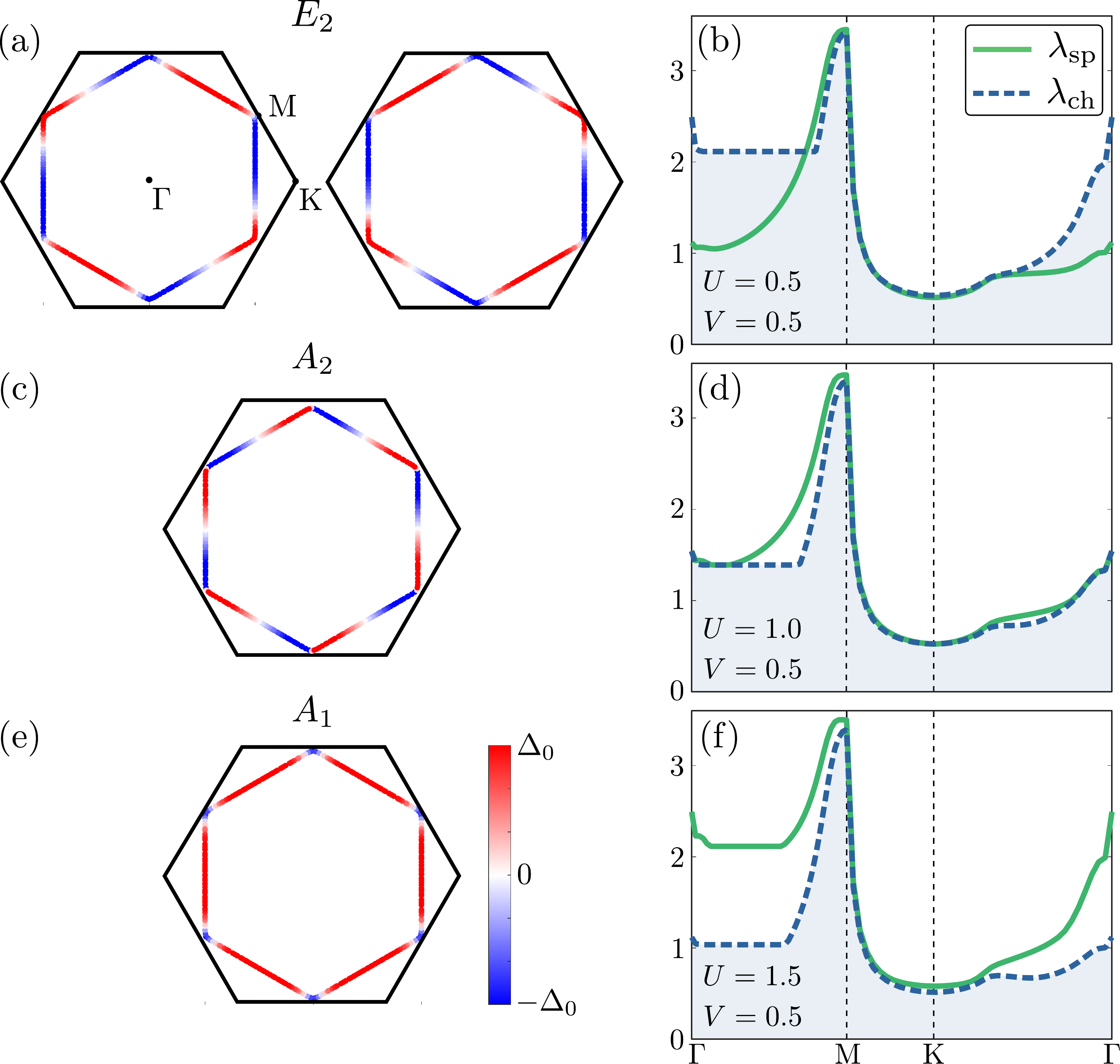}
\caption{Superconducting gap structures displayed on the Fermi surface and corresponding spin and charge susceptibilities very close to the upper van Hove filling, $\mu=0.001$, for $V=0.5$ and increasing values of $U$. The gap structures shown in (a), (c), and (e) correspond to the different solutions shown in Fig.~\ref{fig:phasediagrams}(c) for $V=0.5$ as $U$ is increased. In (a), we show the case $U=0.5$ where we recover the $E_2$ solution, but with a higher order nodal gap structure as compared to the low-$U$ region of Fig.~\ref{fig:phasediagrams}(a). In this case, both spin and charge susceptibilities in (b) exhibit a peak at $M$, and the charge susceptibility also displays a smaller peak at $\Gamma$. As $U$ is increased, the solution changes to the spin-singlet $A_2$ irrep shown in (c) for $U=1.0$. The $A_2$ irrep has 12 symmetry-imposed nodes and the lowest-order lattice harmonic, which is dominant in this case, corresponds to a so-called $i$-wave solution. As seen in (d), both susceptibilities are peaked at $M$ in this case, while the peak near $\Gamma$ is suppressed. In (e), we show the $U=1.5$ case which leads to a nodal spin-singlet $A_1$ solution. While the peak near $M$ remains in both susceptibilities, as shown in (f), the spin susceptibility has now developed a subleading peak near $\Gamma$.}
\label{fig:gapsuppervHu_0001}
\end{figure}

Figures~\ref{fig:phasediagrams}(b) and (c) show the phase diagram obtained by reducing the electron concentration to be even closer to the upper van Hove filling, where the Fermi surface approaches a perfect hexagon. In this case, the corners of the hexagon nearly touches the van Hove points at M, and these singular points strongly dominate the susceptibilities and determine the associated pairing structure. Therefore, is it not surprising that the leading superconducting order changes substantially, as seen by comparing, e.g., Figs.~\ref{fig:phasediagrams}(a) and (c). The rather extreme Fermi surface relevant for Fig.~\ref{fig:phasediagrams}(c) largely removes the triplet regions, and replaces them by nodal $A_{1}$ order, as discussed further below. 

In Fig.~\ref{fig:gapsuppervHd} we provide an overview of the momentum-dependence of the different superconducting gap structures relevant for the leading candidates displayed in Fig.~\ref{fig:phasediagrams}(a). This corresponds to the case near the upper van Hove filling with $\mu=0.08$ and the Fermi surface displayed in the inset of Fig.~\ref{fig:chiRPA}(a). The gap structures shown in Fig.~\ref{fig:gapsuppervHd} are obtained along a line in the phase diagram at $V=0.5$. In the regime $V \gtrsim U/2$ the order parameter belongs either to the spin-singlet $E_2$ irrep or spin-triplet $B_2$ irrep. As seen in Fig.~\ref{fig:gapsuppervHd}(a), the $E_2$ irrep is well-described by the $\{ d_{x^2-y^2}, d_{xy}\}$ lattice harmonics in this case. In the TRSB (chiral) combinations $d_{xy}\pm id_{x^2-y^2}$, this solution thus yields a fully gapped state. In contrast, the $B_2$ solution shown in Fig.~\ref{fig:gapsuppervHd}(b) exhibits symmetry-protected gap nodes. For the cases displayed in Fig.~\ref{fig:gapsuppervHd}(a) and \ref{fig:gapsuppervHd}(b), both the charge- and spin-susceptibilities exhibit pronounced peaks near $M$, as seen in Figs.~\ref{fig:gapsuppervHd}(e) and \ref{fig:gapsuppervHd}(f). However, Fig.~\ref{fig:gapsuppervHd}(f) and (g) highlight a substantial small-$\bf{q}$ weight of the susceptibilities, which we attribute as the origin of the spin-triplet state. 

In the other region, $V \lesssim U/2$, the preferred pairing state for the case shown in Fig.~\ref{fig:phasediagrams}(a) belongs either to the spin-triplet $E_1$ irrep or the spin-singlet $E_2$ irrep, which both appear in highly nodal versions, as seen in Figs.~\ref{fig:gapsuppervHd}(c) and \ref{fig:gapsuppervHd}(d). In other words, both solutions receive substantial contributions from lattice harmonics beyond the leading order. In the spin-triplet state the two components in Fig.~\ref{fig:gapsuppervHd}(c) exhibit coinciding (accidental) nodes and, as a result, any TRSB combination remains nodal. Similarly, the high-$U$ spin-singlet $E_2$ solutions exhibit near-coinciding nodes, as seen in Fig.~\ref{fig:gapsuppervHd}(d). This surprising result arises from the competing nesting vectors shown in panels \ref{fig:gapsuppervHd}(g) and \ref{fig:gapsuppervHd}(h), generating additional accidental nodes. Similar to the origin of the $E_1$ triplet state discussed here, a recent study of finite-$\bf{q}$ triplet pairing with additional nodes was discussed in the context of UTe$_2$~\cite{Kreisel_triplet2022}. For the $E_2$ solutions in this regime, the resulting gap structure exhibits a "coral snake" like pattern as seen from panel Figs.~\ref{fig:gapsuppervHd}(d).  These additional nodes may overlap and generate a nodal, or near-nodal state with very deep minima, TRSB spin-singlet pairing state. The presence of additional non-symmetry-imposed nodes for cases with several competing nesting vectors is similar to results of spin-fluctuation mediated pairing in Sr$_2$RuO$_4$~\cite{Romer2019,Romer2021,Romer2022}. 

The phase diagram for $\mu=0.02$ is shown in Fig.~\ref{fig:phasediagrams}(b). As a result of the near-degeneracy of the pairing eigenvalues depicted in Fig.~\ref{fig:phasediagrams}(e), even a minor change in the electronic filling can have an impact on the resulting pairing instability. Indeed, the $E_1$ triplet region is now dominant when compared to the $B_2$ region present in the $\mu=0.08$ case. Moreover, the $A_2$ solution has appeared for high values of $U$ and $V$. The $E_2$ regions present both in the lower right and upper left parts of Fig.~\ref{fig:phasediagrams}(a) remain in Fig.~\ref{fig:phasediagrams}(b).

Very close to the upper van Hove filling, for $\mu=0.001$, the superconducting phase diagram is displayed in Fig.~\ref{fig:phasediagrams}(c). As mentioned above, despite the marginal change of the electron concentration, the leading superconducting instabilities change substantially from the results shown in Fig.~\ref{fig:phasediagrams}(a) due to the rather extreme nesting of the perfectly hexagonal Fermi surface. The associated gap structures for this case are shown in Fig.~\ref{fig:gapsuppervHu_0001}. In Figs.~\ref{fig:gapsuppervHu_0001}(a), (c), and (e), we show the leading superconducting gap structures for different representative values of $U$, again at constant $V=0.5$. At all interaction strengths, we observe a prominent peak at the $M$ point as shown in Figs.~\ref{fig:gapsuppervHu_0001}(b), (d), and (f). 

In the case of $U=V=0.5$, the M-peak is accompanied by a small-$\bf q$ charge peak. This gives rise to the $E_2$ gap structure shown in Fig.~\ref{fig:gapsuppervHu_0001}(a), resembling the findings of Fig.~\ref{fig:gapsuppervHd}(a), but with a higher order nodal gap structure in the present case. Increasing the on-site interaction strength $U$ enhances the peak at $M$ compared to the small-$\bf q$ peak as evidenced from Fig.~\ref{fig:gapsuppervHu_0001}(d). This promotes the singlet solution $A_2$ displaying a sign change for Fermi surface points connected by $M$ and bears resemblance to the scenario found at the lower van Hove filling presented in Sec.~\ref{sec:sc_lower_vh} below.

A further increase of $U$ boosts a $\bf q \simeq$ 0 peak structure in the spin channel, as seen from the green curve in Fig.~\ref{fig:gapsuppervHu_0001}(f). 
In this rather extreme case, a nodal $A_1$ state is preferred,  which is characterized by opposite gap signs at the van Hove points at $M$ compared to the remaining Fermi surface points.
This structure takes advantage of the small-$\bf q$ structure in the susceptibility and thus the pairing kernel. At the same time, the gap structure avoids a node directly at the $M$ points where the density of states is maximum, contrary to the nodes displayed by the $A_2$ gap solution.

\subsection{Leading superconducting solutions near the lower van Hove filling}\label{sec:sc_lower_vh}

\begin{figure}
 \centering
   	\includegraphics[angle=0,width=\columnwidth]{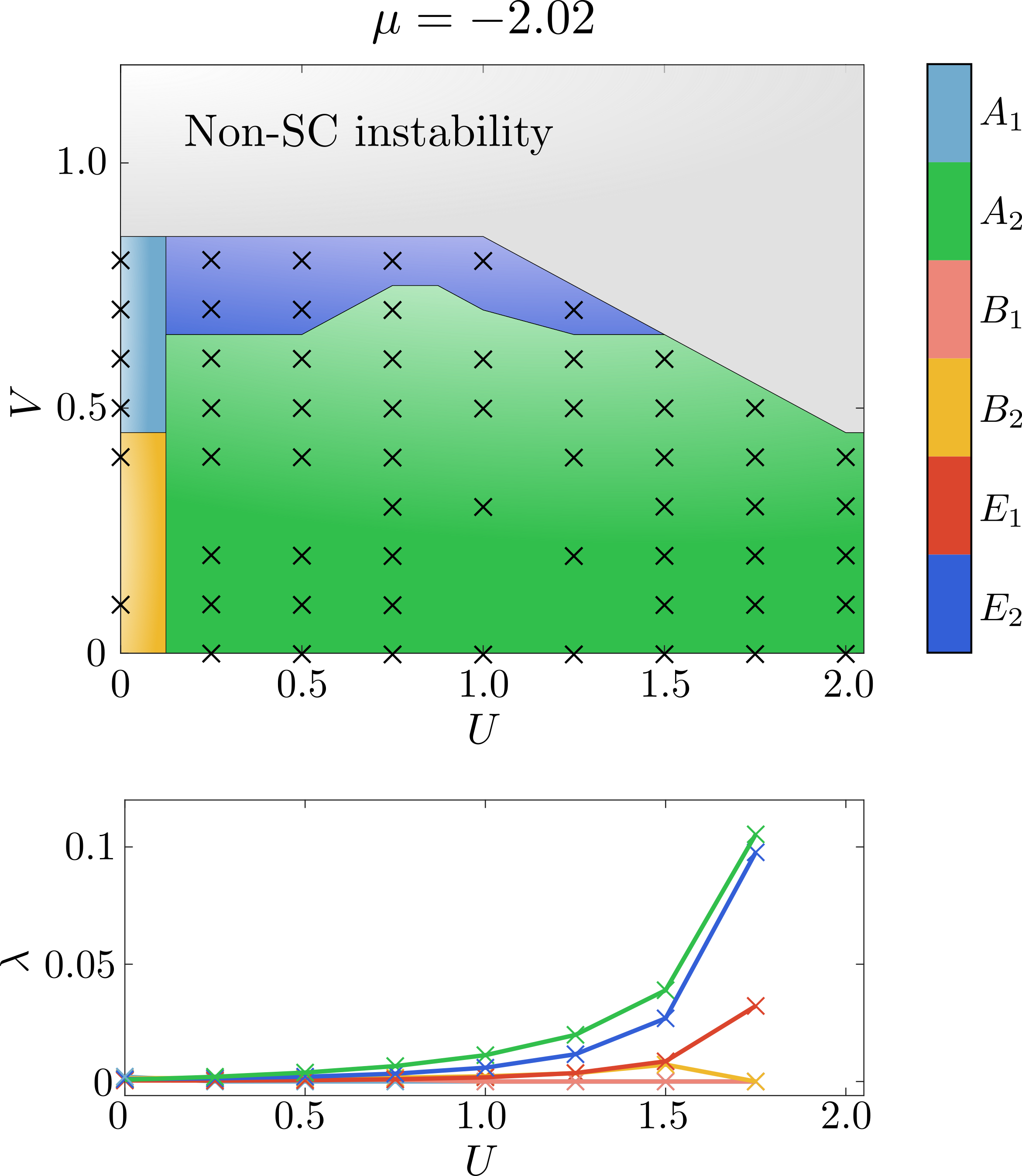}
\caption{(a) Phase diagram showing the leading superconducting pairing symmetry versus onsite interactions, $U$, and NN interactions, $V$, close to the lower van Hove filling with $\mu=-2.02$ corresponding to electron filling $n=1/4-0.006$. The phase diagram is dominated by the spin-singlet $A_2$ solution. (b) Eigenvalues of the linearized gap equation as a function of $U$ for $V=0.5$.}
\label{fig:phasediagramlower}
\end{figure}

\begin{figure}
 \centering
   	\includegraphics[angle=0,width=\linewidth]{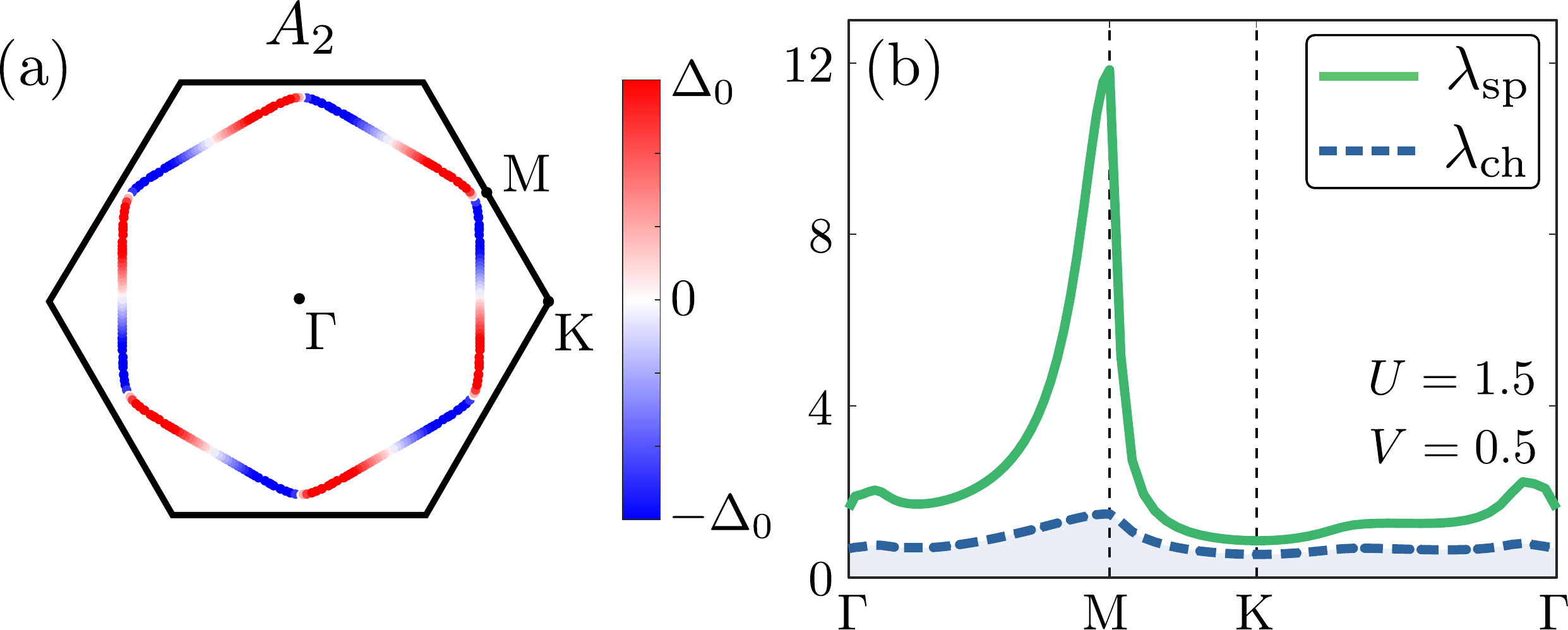}
\caption{(a) Superconducting spin-singlet $A_2$ gap structure on the Fermi surface very close to the lower van Hove filling, $\mu=-2.02$, for $V=0.5$ and $U=1.5$. The dominance of the $A_2$ solution is attributed to the robust peak in the susceptibility near $M$ shown in (b).}
\label{fig:gapslowervH}
\end{figure}

As illustrated in Fig.~\ref{fig:chiRPA}, the susceptibilities are strongly affected by the sublattice distribution along the Fermi surface. To illustrate the impact of this property on the superconducting gap structure, we solve the linearized gap equation also near the lower van Hove point, with $\mu=-2.02$, corresponding to a filling of $n=1/4-0.006$. The associated Fermi surface and sublattice distribution is shown in Fig.~\ref{fig:chiRPA}(d). The phase diagram near the lower van Hove filling is shown in Fig.~\ref{fig:phasediagramlower}(a). In contrast to the cases discussed in Sec.~\ref{sec:sc_upper_vh}, the leading superconducting pairing is dominated by the spin-singlet $A_2$ irrep. At this filling, the pairing is ``non-frustrated'' as indicated by the relatively unchanging phase diagram in Fig.~\ref{fig:phasediagramlower}(a) and the well-separated, monotonously increasing eigenvalues in Fig.~\ref{fig:phasediagramlower}(b). The detailed gap structure of the spin-singlet $A_2$ irrep is shown in Figs.~\ref{fig:gapslowervH}(a). Throughout the phase diagram, this gap structure appears free of additional accidental nodes, and exhibits only the 12 nodes imposed by symmetry. This is distinct from the situation near the upper vHS, where large regions of the phase diagrams were dominated by solutions with a large number of accidental nodes. We attribute the dominance of the $A_2$ irrep to the presence of a single peak in the susceptibility, as seen in Figs.~\ref{fig:gapslowervH}(b). The $A_2$ solution is the only allowed spin-singlet gap structure with opposite signs of the gap on all parallel Fermi surface segments. Finally, we note that the sign-changes of the $A_2$ gap function in Fig.~\ref{fig:gapslowervH}(a) are very steep near the van Hove points. This is a consequence of two competing effects. On the one hand, an effective attraction in the singlet channel is achieved by the order parameter having opposite signs on nested segments of the Fermi surface. On the other hand, the density of states diverges at the van Hove points, implying that the gain in condensation energy is largest if these points are gapped out. This balancing act results in a steep gap function near the $M$ points.

\section{Discussion}

We have solved the linearized BCS gap equation relevant for the simple kagome lattice with superconducting pairing generated by spin and charge fluctuations. In this framework we included both on-site $U$ and nearest-neighbor $V$ interactions within a microscopic RPA approach keeping all bubble and ladder diagrams entering the pairing vertex. When $V$ is only included through the bubble diagrams as assumed in Ref.~\cite{Wu2021}, the formalism simplifies substantially. While this is enough to capture the physics of the CDW instability, we find that this approximation tends to slightly overestimate the role of spin-fluctuations as compared to the case where $U$ and $V$ processes are included in both bubble and ladder diagrams.

Several earlier theory works studied superconducting pairing in the kagome Hubbard model using a variety of methods, and found leading chiral $d_{x^2-y^2} \pm i d_{xy} $ superconducting order near the upper van Hove filling~\cite{Kiesel2012,Yu2012,Wang2013}. In addition, it has been pointed out that also $f$-wave triplet channels or $s$-wave superconductivity can be competitive near this electron filling~\cite{Kiesel2012,Kiesel2013,Wu2021,Wen2022,Tazai2022}. Overall, these results are in agreement with the findings reported in this work, even though direct comparison is difficult due to different approximations or input band structures. Here, we have provided a comprehensive study of the ``pairing landscape'' for the simple one-orbital kagome lattice as a function of electron filling and interaction strengths. Near the upper van Hove point, in the physically most reasonable regime where $U \gtrsim 2V$, we find three dominant pairing instabilities depending on the precise electron density and the values of $U$ and $V$. As seen in Fig.~\ref{fig:phasediagrams}, these are the nodal $A_1$ spin-singlet, the spin-triplet $E_1$ irrep, and the spin-singlet $E_2$ irrep. The latter two lead to gap structures that are highly oscillatory on the Fermi surface, as seen in Figs.~\ref{fig:gapsuppervHd}(c) and \ref{fig:gapsuppervHd}(d), and both are expected to break time-reversal symmetry.

One of the main results of our calculations is the near-degeneracy of several symmetry-distinct pairing states near the upper van Hove filling, as evident in Fig.~\ref{fig:phasediagrams}. This is related to the sublattice structure of the electronic states, which generates several competing susceptibility contributions, and is quite different from, e.g., high-$T_c$ cuprates or iron-pnictide superconductors where similar calculations strongly favor a $d_{x^2-y^2}$-wave or $s_\pm$ pairing structure, respectively~\cite{Romer2015,HIRSCHFELD2016197}. This near-degeneracy means that additional effects -- e.g., self-energy corrections, details of the band-structure, spin-orbit coupling, and further longer-ranged interactions -- not included in the present formalism may tip the balance and alter the hierarchy of the leading pairing states~\cite{Kreisel_2017,Romer2019,Bjornson_PRB2021,Kreisel_PRL2022}. Thus, to address particular compounds such as $A$V$_3$Sb$_5$, one needs to consider material-specific modelling in order to accurately pinpoint the leading theoretical superconducting ground state. However, even in that case, the near-degeneracy is expected to remain due to the sublattice interference near the upper van Hove point~\cite{Kiesel2012,Kang2022Twofold}. Such near-degeneracy may imply that the superconducting state in actual compounds is sensitive to external parameters such as sample preparation and easily tunable by, e.g., pressure. There are indications that this may apply to the $A$V$_3$Sb$_5$ compounds~\cite{Wang_triplet,Guguchia2022Tunable}.

As discussed in the Introduction, the current status of the superconducting gap structure in the $A$V$_3$Sb$_5$ kagome metals is controversial with reports of both nodeless superconductivity and nodal quasiparticles~\cite{Ortiz2021Superconductivity,Duan2021,Gupta2022,Roppongi2022,Xu_PRL_2021,Chen2021Roton,Liang2021,Zhao_2021}. In fact, for these compounds it is still being debated whether the mechanism is indeed unconventional, or if superconducting pairing could be mediated by phonons~\cite{Tan2021,Zhong2022}. A substantial fraction of the current experiments on $A$V$_3$Sb$_5$ points against true nodes and spin-triplet Cooper pairs~\cite{Ortiz2021Superconductivity,Duan2021,Gupta2022,Roppongi2022,Xu_PRL_2021,Mu_2021}. In this respect, from the present calculations the absence of nodal quasiparticles at the lowest temperatures points against the $B_2$ and $E_1$ triplet states. Both these states could remain close-by however, which may explain a spin-triplet signature in certain experiments~\cite{Wang_triplet}, even if triplet order is not the preferred ground state. On the other hand, for the leading spin-singlet orders found in this work, the presence or absence of nodal low-energy quasiparticles does not uniquely identify the preferred pairing state. Focusing again on the regime $U \gtrsim 2V$, the $A_1$ state discussed in Sec.~\ref{sec:sc_upper_vh}, and shown in Fig.~\ref{fig:gapsuppervHu_0001}(e), exhibits true nodes. On the other hand, the $E_2$ chiral TRSB state in this parameter regime features either deep minima or accidental nodes, contrary to conventional expectations from lowest-order $d_{xy}/d_{x^2-y^2}$ harmonics. Thus, the $E_2$ state currently stands out as the most likely candidate for the (unconventional) superconducting order parameter relevant for the $A$V$_3$Sb$_5$ superconductors. This state breaks time-reversal symmetry, is of spin-singlet nature, and possesses an anisotropic full gap with deep minima. However, additional experiments and material-specific theoretical modelling are required for reaching consensus. In this respect, an important open question for the $A$V$_3$Sb$_5$ materials is the impact of the CDW on the superconducting order. Feedback effects from the charge order are not included in the presented framework, and constitute an interesting open topic for future exploration~\cite{Lin2021_Multidome,Guguchia2022Tunable}. On the other hand, the Fermi surface reconstruction induced by the CDW is weak~\cite{Li2021Spectroscopic}, and the minimal one-orbital model may still provide useful guidance in this case. However, the opening of a CDW-induced gap at the van Hove points~\cite{Kang2022Twofold} would impact the relative importance of these points in the gap equation.

The CDW phase appears to compete with superconductivity, and the two have a dichotomous dependence on, e.g., pressure or doping~\cite{Chen2021Double,Guguchia2022Tunable,Oey2022Fermi,Gupta2022Two}. As a consequence, our results can apply to the pressurized or doped compounds, for which no CDW phase has been observed~\cite{Chen2021Double,Guguchia2022Tunable,Oey2022Fermi,Gupta2022Two}. Indeed, the report of a TRSB nodeless gap with deep minima in pressurized KV$_3$Sb$_5$ and RbV$_3$Sb$_5$ appears consistent with the $E_2$ gap structure of Fig.~\ref{fig:gapsuppervHd}(d). Here, the several near-accidental nodes can lead to deep minima in the gap, but the Fermi surface remains nodeless. While it is tempting to draw similar conclusions for LaRu$_3$Si$_2$, YRu$_3$Si$_2$, and LaIr$_3$Ga$_2$, to our knowledge, the Fermi surfaces for these compounds have not yet been experimentally measured. Hence, it is unclear to what extent the one-orbital model is applicable to these materials.

\section{Conclusions}

In summary, we have mapped out the superconducting phase diagram near the two van Hove fillings of the simple kagome lattice within spin- and charge-fluctuation mediated superconductivity based on on-site and nearest-neighbor repulsive interactions. In addition, we have surveyed the resulting gap structures and their dependence on electron filling and interaction parameters. The hierarchy of the Cooper pairing is highly frustrated at the upper van Hove filling due to the competition between several nesting vectors in the susceptibilities. For the same reason, the momentum structure of the pair potentials typically exhibits several accidental nodes in addition to the symmetry-imposed ones. Our calculations find both singlet and triplet orders as possible relevant pairing solutions. However, given the bulk of current experiments on the AV$_3$Sb$_5$ compounds, a near-nodal version of the TRSB spin-singlet $E_2$ solution stands out as a leading candidate for unconventional superconductivity in these materials. 

\begin{acknowledgments} We acknowledge helpful discussions with E. Hellebek and A. Kreisel. A.T.R. and B.M.A. acknowledge support from the Independent Research Fund Denmark grant number 8021-00047B. S.B. and R.V. acknowledge support from the Deutsche Forschungsgemeinschaft
(DFG, German Research Foundation) through QUAST
 FOR 5249-449872909 (Project P4). M.H.C. has received funding from the European Union’s Horizon 2020 research and innovation
programme under the Marie Sklodowska-Curie grant agreement No 101024210.
\end{acknowledgments}

\bibliography{kagome_SC}

\end{document}